\documentclass[showpacs,amsmath,amssymb,floatfix,prl,twocolumn]{revtex4}
\usepackage{graphicx,amsmath}

\begin{document}

\title {Phonon assisted dynamical Coulomb  blockade in a thin suspended graphite sheet.}
\author{A. Chepelianskii}
\affiliation{Univ. Paris-Sud, CNRS, UMR 8502, F-91405 Orsay Cedex, France}
\author{P.Delplace}
\affiliation{Univ. Paris-Sud, CNRS, UMR 8502, F-91405 Orsay Cedex, France}
\author{A.Shailos}
\affiliation{Univ. Paris-Sud, CNRS, UMR 8502, F-91405 Orsay Cedex, France}
\author{A.Kasumov}
\affiliation{Univ. Paris-Sud, CNRS, UMR 8502, F-91405 Orsay Cedex, France}
\author{R.Deblock}
\affiliation{Univ. Paris-Sud, CNRS, UMR 8502, F-91405 Orsay Cedex, France}
\author{M.Monteverde}
\affiliation{Univ. Paris-Sud, CNRS, UMR 8502, F-91405 Orsay Cedex, France}
\author{M.Ferrier}
\affiliation{Univ. Paris-Sud, CNRS, UMR 8502, F-91405 Orsay Cedex, France}
\author{S.Gu\'eron}
\affiliation{Univ. Paris-Sud, CNRS, UMR 8502, F-91405 Orsay Cedex, France}
\author{H.Bouchiat}
\affiliation{Univ. Paris-Sud, CNRS, UMR 8502, F-91405 Orsay Cedex, France}

\pacs{63.22.Np, 73.23.Hk, 73.21.Ac} 
\begin{abstract}
The differential conductance  in a suspended few layered  graphene sample is found to exhibit a series of  quasi-periodic  sharp dips as a function of bias at low temperature. We show that they can be understood within a simple model of dynamical Coulomb blockade  where  energy exchanges  take place between the  charge carriers transmitted trough the sample and a dissipative  electromagnetic environment with a resonant phonon mode strongly coupled to the electrons. 
\end{abstract}

\maketitle
One of the great challenges of molecular electronics is to access
electron-phonon coupling at the single molecule level.  Mechanically tunable atomic break junctions  with trapped  small molecules  such as ($H_2$, $D_2$, $H_20$) have been shown to  exhibit a spectroscopic signature of their characteristic  phonon modes \cite{VanRuyten}.  The signature of phonons is  also  spectacular in the Coulomb blockade regime for a  molecular single electron transistor: The typical resonant tunneling peaks as a function of gate or source-drain voltage are surrounded by satellites, which correspond to the emission or absorption of one or several phonons. Specific vibrational modes were identified in this way in fullerenes and suspended carbon nanotubes \cite{park,dekkervib,vanderzant}. Theoretical models \cite{flensberg03a,flensberg03b,mitra04}  were developed  to describe these vibrational side bands in molecular  transistors, involving either a quantum or classical treatment of  the electron phonon coupling. In all these investigations the single level spacing within the molecule is larger than the energy of the vibration modes coupled to the molecule, so that only a single molecular level needs to be considered. In the present work we investigate the opposite limit of a mesoscopic dot where  both the single level spacing and Coulomb charging energy  are smaller than the energy of the phonon mode considered.  Moreover the  transmission of the electrodes corresponds to an intermediate  tunneling regime described by the physics of dynamical Coulomb blockade (DCB).   The samples  are  micron size few  atomic layer  graphite foils suspended between two platinum  electrodes. The differential conductance exhibits at low temperature  around zero bias a power law increase   characteristic of  DCB  through the  contacts. The graphite  foil sample itself constitutes  the  electromagnetic environment. More original, on the thinest sample (with 30 graphene layers) a series of periodic replica of this Coulomb  blockade anomaly was detected at multiples of 20 meV, corresponding   approximately to the   lowest energy out of plane optical mode in graphite (ZO')\cite{rubio}.  These  sharp dips  were not observed on  two control   graphite samples which were likewise suspended, had similar resistance and lateral dimensions  but were  more than 30 times thicker.  We analyse these results with a simple model,  inspired by \cite{mitra04}, of a mesoscopic island connected  to  electrodes via  tunnel barriers.   We model the island   by a  continuous  electronic spectrum   coupled to  a single phonon mode, which leads to an oscillating   transmission of the barriers at the contacts \cite{mitra04}. This model can  also be solved using the so called   $P(E)$ theory developed by Ingold and Nazarov  \cite{ingold}  to describe energy exchanges of a Coulomb blocked tunnel junction with a dissipative electromagnetic environment and presents a striking analogy with the behavior of a tunnel junction coupled to an electromagnetic resonator in series with an ohmic environment as investigated by \cite{devoret95}. We finally deduce from  the field dependence of these dips an order of magnitude for the  relevant electron-phonon coupling parameter in the system.   

\begin{figure}
\begin{center}
\includegraphics[clip=true,width=9cm]{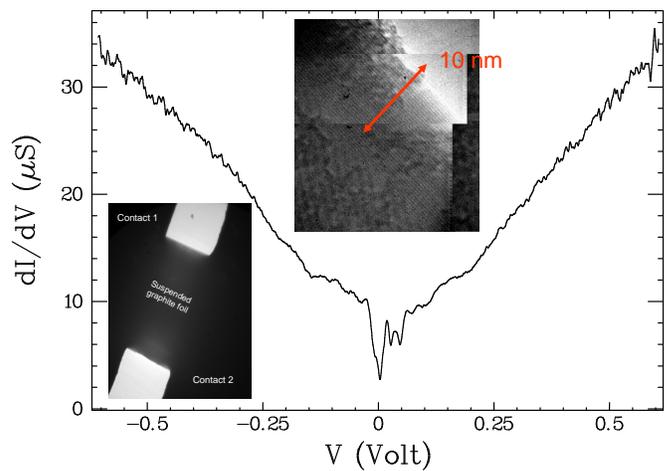}
%\leavevmode
%\epsfxsize=8cm
\caption{ Bias dependence of the differential conductance measured on a suspended thin foil of graphite. Inset: a) transmission electron microscopy picture of the sample. b) Side view taken at high resolution from which it is possible to estimate the number of  graphene layers of the order of 30.
\label{samplepanddIdVhighV}}
\end{center}
\end{figure}

	The samples were prepared by exfoliation of a highly oriented pyrolytic graphite (HOPG) single crystal and deposition across a slit etched in a silicon nitride membrane separating two Pt metallic contacts. The number of graphene layers  was estimated from transmission electron microscopy pictures, see fig.\ref{samplepanddIdVhighV} for the thinnest sample, which contains between 25 and 35 layers. The electrical contacts were obtained just by  pressing  the sample onto the electrodes. Thus the sample resistance, $ R_t= 100 k\Omega$ at 4.2 K, mainly consists of the resistance at the contact, and increases as the temperature is reduced.  The square 
resistance  of the graphite layer  itself is not expected to be larger than $5 k\Omega$, the maximal resistance  of a single graphene sheet. 
The differential conductance  $dI/dV$ was  either measured by    modulating the voltage bias  and measuring the current modulation by  standard lock-in detection  or  deduced from the differential resistance obtained by applying a small ac  current  of typical amplitude 1 to 10nA superimposed to a dc  current bias. The dc voltage drop  V through the sample was then  deduced by integration of $dV/dI =f(I)$.
The triangular shape of $dI/dV =f(V)$ observed at high bias (above 0.15 V,  see fig.\ref{samplepanddIdVhighV}) can be related to the  linear dependence of the density of states  $\nu(E$),  characteristic of the band structure of graphene as well  as of graphite at high enough energy \cite{bandstructurewallace}: Indeed, the electronic transmission  between the graphite sample  and the underlying  electrodes is low, so that the voltage drop occurs mainly   at the contacts.  The differential conductance can then be written as:
 
 \begin{equation}
 dI/dV \propto  \Gamma_L \Gamma_R( \alpha \nu(EF +\alpha eV) +  (1-\alpha) \nu (EF - (1-\alpha)eV))
 \label{asym}
 \end{equation}
 
where  $\Gamma_L$ and $\Gamma_R$ are the transmissions of the left and right contacts respectively. The parameter $1/2 \leq \alpha \leq 1$  characterizes  the asymmetry of the contact resistances and voltage drop,  $\alpha = 1/2$ corresponds to symmetrical contacts  with $V/2$   voltage drop  at each contact. The asymmetry observed between positive and negative bias  is attributed to a combination of a slight doping  of the sample together with some asymmetry in the transmission of the electrodes.

\begin{figure}
\begin{center}
\includegraphics[clip=true,width=9cm]{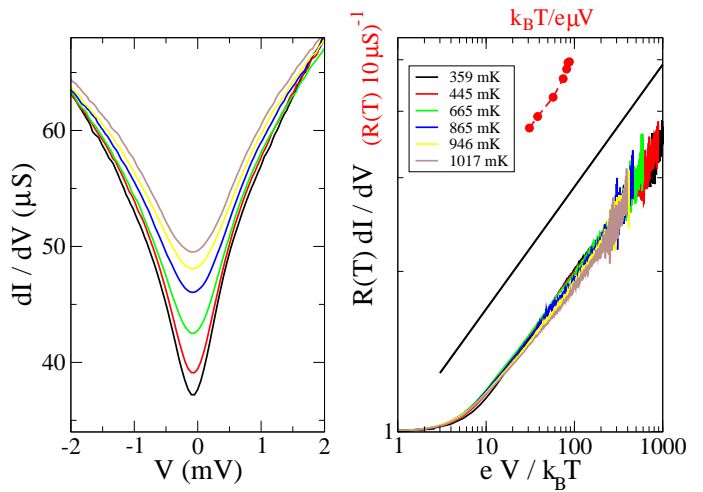}
%\leavevmode
%\epsfxsize=8cm
\caption{Differential conductance in the vicinity of zero bias measured on a thick suspended graphite sample  measured at several temperatures  between 300mK  (lower curve) and 1K (upper curve) The continuous line  is a power law of exponent 0.25. The data can be rescaled according to  eq.\ref{scaling} with z =0.25.  Full circles: temperature dependence of the zero bias resistance.
\label{scalingdynamicCB}}
\end{center}
\end{figure}

 We now focus on the conductance at low voltage (below 0.12 V) which exhibits a  pronounced dip at zero bias. This behavior   is  characteristic of   Coulomb blockade  through a small capacitance tunnel junction   in series with  a dissipative  electromagnetic environment  which  can  exchange energy with  the tunneling quasi particles on a scale much smaller than the charging energy.   This yields  the so called Dynamical Coulomb  Blockade (DCB)  \cite{ingold}. The differential conductance data is  expected to follow  a scaling behavior as a function of bias and temperature:  
 \begin{equation}
   G(V) = dI/dV = T^z f(eV/k_BT)
   \label{scaling}
 \end{equation}  
 
with $\lim_{x\rightarrow 0} f(x)= Cst$ and $\lim_{x\rightarrow \infty} f(x)=x ^z$, and the exponent $z$  is expected to be $\alpha^2 R/R_Q$  inthe case of Two asymmetric junctions, where R is the resistance of the environment and $R_Q= h/2 e^2$ is the  resistance quantum.   The data shown in fig.\ref{scalingdynamicCB}  yields z of the order of $0.25\pm 0.05 $. Such a power law dependence  also was found in the two other thicker samples, with similar exponents. It is thus reasonable to assume that the dissipative  ohmic environment in the present  case is constituted by the  few top graphene layers of the graphite samples.  Note  also that a similar behavior was already observed  on multiwall carbon nanotubes  with low conductance contacts \cite{bachtold}. 

\begin{figure}
\begin{center} % fit with theory.
\includegraphics[clip=true,width=9cm]{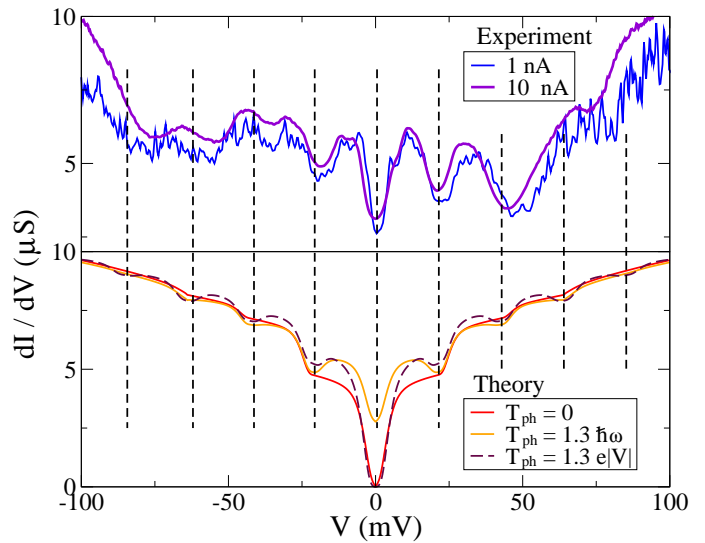}
\leavevmode
\caption{ (a)Differential conductance measured at 70mK on the thin graphene foil depicted in fig.\ref{ssamplepanddIdVhighV},  for to current excitations 1nA (thin line) and 10nA (bold line). Note the  sharp dips nearly equally spaced  by 20mV , except the  broad dip  centered  around 50mV   which can be decomposed into 2 overlapping dips centered around 40  and 60mV as suggested by the data taken at 1nA ac excitation.(b)  Theoretical fit for different phonon temperatures  with  the following parameters deduced from the geometry of the sample and the differential conductance data at low bias described by DCB : $\alpha = 0.75$, $R_T = 125 {\rm k \Omega}$,  $\Delta = 200 {\rm mV}$,  $e^2/C = \hbar \omega / 4$. The environment is described with a resistance  $R / R_Q = 1$.  The only free adjustable parameter is $\lambda = 0.7$.
\label{peaksg}}
\end{center}
\end{figure}

More original,  as shown on  fig. \ref{peaksg}, is the  bias dependence  of the differential conductance   measured on  the thinest sample investigated which is 10 nm thick and contains thirty graphene layers.  It exhibits a series of eight sharp  dips  resembling the zero bias one and nearly equally spaced   by $20\pm 2 mV$. Their  amplitude  decreases with increasing voltage except for the   broader dip  at 50 mV which can be decomposed into 2 overlapping negative peaks centered around 40  and 60mV as suggested by the data taken at 1nA ac excitation.  The energy scale  of 20 meV does not correspond to any simple electronic energy scale in the sample, whose charging energy is in the meV range and level spacing  in the $10\mu eV$ range. On the other hand the  lowest energy  optical phonon in graphite (so called Z0') has an energy of 15meV  \cite{rubio}. This mode, which emerges from the out of plane transverse acoustic mode of graphene  \cite{rubio, xray}, is only present in graphite and  corresponds to the two neighboring, non equivalent, planes vibrating in phase opposition   along the c axis. This phonon mode has been observed experimentally in graphite by inelastic X-ray scattering \cite{xray} and scanning tunneling spectroscopy \cite{vitali} with an energy of $15 \pm 1meV$.
The observed   peak  positions at multiple values of 20 mV  instead of 15 mV can be attributed to the asymmetry of the contacts which  corresponds     to  the parameter $\alpha \simeq 0.75$ in eq.\ref{asym}. These dips are only observed on the 10 nm thick  foil and were not  detected on the two thicker (more than 100 nm) samples. This can be understood considering  that the  conversion from electric energy (depending only on the resistance of the tunneling barriers) into mechanical vibrations   leads to an   induced vibration amplitude inversely proportional to the number of layers in the graphite foil.  The suspended character of the sample is also  essential, since interaction with a substrate 
 suppresses considerably  the amplitude of induced vibrations as already demonstrated on carbon nanotubes \cite{dekkervib}.  Note  that STM spectroscopy on bulk graphite \cite{vitali} also reveals inelastic contributions due to plasmons which are not detected in the present experiment.

 In order to explain the data  more quantitatively we  extended   the work of Mitra et al. \cite{mitra04} on the phonon assisted Coulomb staircase observed in the transport through fullerenes molecules. The coupling between the ZO'-phonon mode and the electrons in the graphite sample is described using a Holstein Hamiltonian \cite{Holstein}. 
In the  absence of disorder this Hamiltonian $H_G$ reads: 
\begin{align}
H_{G} = \sum_k \epsilon_k c_k^+ c_k + \lambda \; \hbar \omega \sum_k c_k^+ c_k (b^+ + b) + \hbar \omega \; b^+ b
\end{align}
where $c^+_k$ and $c_k$ are the fermionic electron creation and annihilation operators 
in momentum space, $\epsilon_k$ the electronic energy, 
and $b^+$ and $b$ the  creation and annihilation operators of the bosonic ZO'-phonon of frequency $\omega$. In contrast with  previous work \cite{mitra04},  the electronic energy level spacing is small compared to the phonon energy $\hbar \omega$. The  parameter $\lambda $ is the  dimensionless electron phonon coupling constant which we assume  to be of the order of unity  like in carbon nanotubes \cite{lambda}.
The coupling to the leads is then described in a dynamical Coulomb-blockade formalism \cite{ingold}, 
by a Hamiltonian of the form $H_T = \sum_{k,k'} T_{k,k'} a_k c^+_{k'} e^{-i \phi} + hc$.
Here $a_k$ are the electron annihilation operators in the leads, $T_{k,k'}$ are 
the tunnel amplitudes and $\phi$ is a phase operator describing the electromagnetic environment 
of the junction. 

The Hamiltonian $H_G$ can be diagonalized with a canonical Lang-Firsov transformation: 
$b' = e^{-S} b e^{S}$, $H_G' = e^{-S} H_G e^{S}$ where $S = \lambda \sum_k c_k^+ c_k (b - b^+)$. In the limit when the 
charging energy of the sample is negligible \cite{mitra04} the transformed Hamiltonian reads simply 
$H_G = \sum_k  \epsilon_k c_k^+ c_k + \hbar \omega \; b^+ b$ where we have omitted the primes for the 
transformed operators. In the transformed basis the transfer Hamiltonian $H_T$ is given by :
\begin{align}
H_T = \sum_{k,k'} T_{k,k'} a_k c^+_{k'} e^{-i \phi + \lambda (b - b^+)} + hc 
\label{eqeph}
\end{align}
This expression is obtained by expanding the product $e^{-S} H_T e^{S}$ under the assumption that the 
environment phase $\phi$ commutes with the phonon operators. It shows that the coupling to phonons essentially 
changes the phase operator of the junction. As a result the current through the junction 
can be expressed with  an effective $P(E)$ function describing the probability 
of electrons to loose an energy $E$ in a tunnel transition as in usual DCB theory. 
Since the  electromagnetic environment and phonon operators commute, this function can be expressed as a 
convolution 
\begin{align}
P(E) = \int d E' P_{env}(E') P_{ph}(E - E')
\label{PE}
\end{align} where $P_{env}(E)$ is 
the probability of emitting a photon of energy $E$ in the $RC$ environment of the junction 
and $P_{ph}$ is the probability of emitting a phonon in the sample. The probability distribution 
$P_{ph}$ can be obtained by noticing that the corresponding phase operator $i \lambda (b - b^+)$ is 
analogous to that of an electromagnetic $LC$ circuit with resonant frequency $\frac{1}{\sqrt{LC}} = \omega$ \cite{ingold}
(this result can also be obtained directly by tracing out the phonon degrees of freedom in eq.\ref{eqeph}): 
\begin{align}
P_{ph}(E) = e^{-\lambda2 \coth( \frac{\beta \hbar \omega}{2} )} \sum_k \delta( E - k \hbar \omega_0 ) e^{k \beta \hbar \omega / 2} I_k( \frac{\lambda2}{ \sinh( \beta \hbar \omega / 2 ) } ) 
\label{PEPH}
\end{align}
where $\beta$ is the inverse of the phonon temperature $T_{ph}$. Using Eqs.~(\ref{PE} and \ref{PEPH}) 
and standard expressions for current as a function of $P(E)$ \cite{ingold} it is possible 
to compute the current through the sample. 
For example in the case of symmetric contacts ($\alpha = 1/2$) and constant density of states, 
$d I/ d V = \frac{1}{R_T} \int dE \left( P(e V / 2 - E) + P(-e V / 2 - E) \right)$ 
where $R_T $ is the tunnel resistance  of each contact.  Note the similarity with the DCB in a tunnel junction in series with a LC resonator \cite{devoret95}. The case of an energy  dependent density of states , 
and of asymmetric contacts can readily be included by straightforward
generalization of this expression  according to eq.\ref{asym}. For comparison with the experiments we assume that the density of states of graphite  is of the form: 
$\nu(E) = \nu_0 ( 1 + \frac{|E|}{\Delta} )$. This formula is exact for bilayer graphene with $ \Delta \approx 400 meV $ \cite{Bilayer} and $|E|< \Delta $. 
We also take the values of $R_T$, environment resistance $R$ and charging energy  deduced from the geometry and conductance data at low bias. The only free adjustable parameter is $\lambda = 0.7$. As shown on fig.\ref{peaksg} the agreement with experimental data is only qualitative especially at zero temperature , where theory predicts  zero conductance at zero bias which is not observed experimentally . Better agreement is found when a  finite phonon temperature is introduced. Since the sample is suspended the populations of phonons are supposed to be strongly out of equilibrium. We have tentatively introduced a phonon temperature increasing linearly with bias which leads to  better agreement with experimental data. However, surprisingly, the best fit is obtained by imposing a bias-independent phonon temperature which does not seem a priori  very physical. We have also in our model neglected the electronic temperature which is expected to be in the K range. Moreover we have not included the expected broadening of the phonon modes (even at very low temperature) due to the strong coupling to electrons. This may explain why the best fit corresponds to a  finite phonon  temperature.
\begin{figure}
\begin{center}
\includegraphics[clip=true,width=9cm]{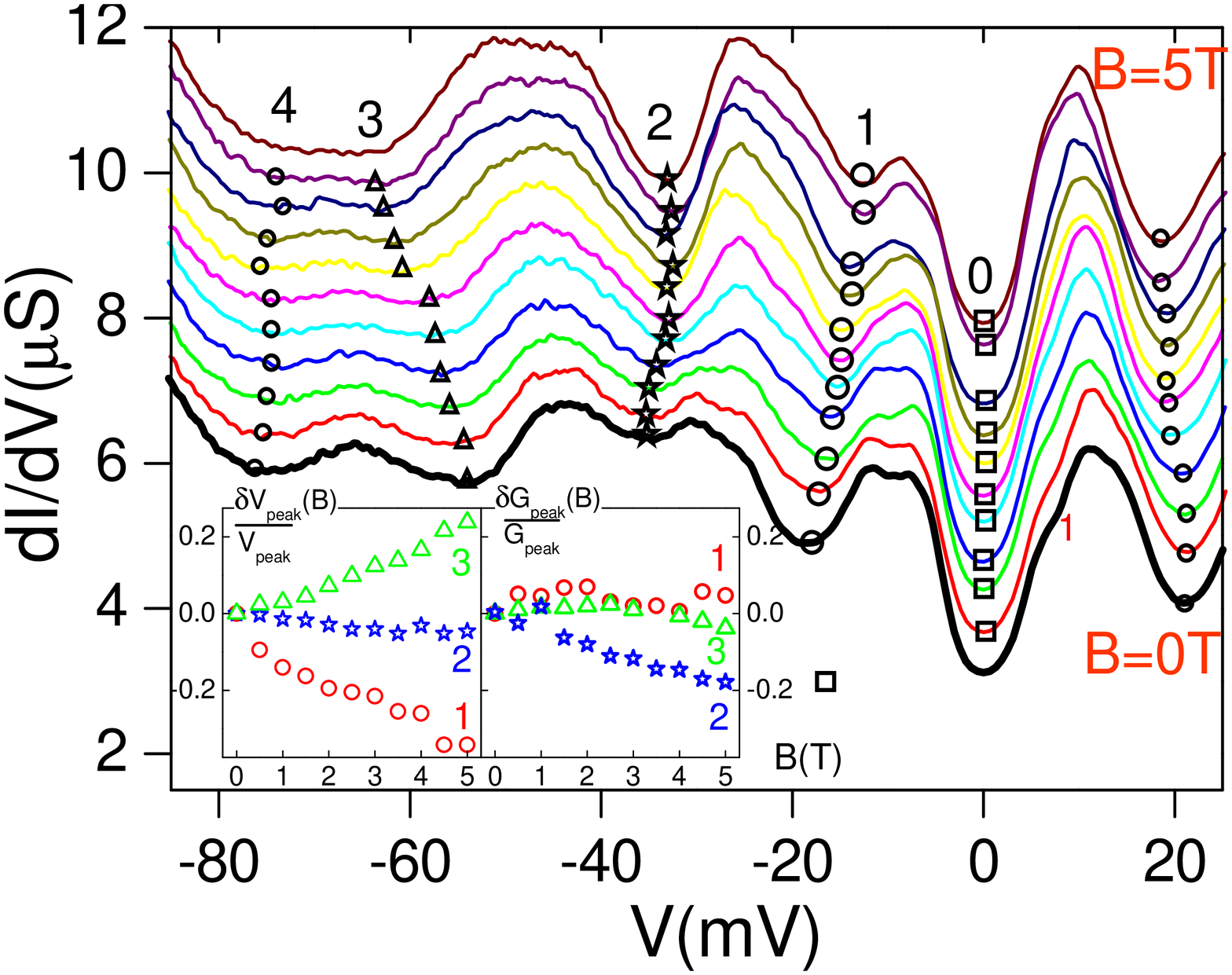}
%\leavevmode
%\epsfxsize=8cm
\caption{Evolution of the differential  conductance dips  with magnetic field. Insets: field dependence of the amplitude and position of the dips labeled 1,2 and 3.
\label{magneticfield}}
\end{center}
\end{figure}

In the following we discuss the magnetic field dependence of the dips in the differential conductance   ( fig. \ref{magneticfield}). They vary both in amplitude and position between 0 and 5 T . Whereas the first two dips   and the forth one are shifted toward lower frequency  with increasing magnetic field (as seen both at positive and negative
bias) the third peak is shifted toward high frequency  (we consider here the negative bias data since the second and third dips at positive bias can barely  be resolved as discussed above).   The amplitude of the dips (see inset in fig.\ref{magneticfield}) may decrease or increase with magnetic field (peak 2 and 3), or vary in a non monotonic way (peak 0 and 1).  The relative shifts of the dips with magnetic field by an typical amount of 5 to 20$\%$ are of the same order of magnitude as the  relative variations of their  magnitudes. We attribute   these effects   to the   field dependent density of states of  the graphite foil in the field range where Shubnikov de Haas oscillations just start to show up. Even if these observations are not yet understood in detail they indicate a strong electron phonon coupling and justify the value of
$\lambda = 0.7$ since  a relationship such as $\Delta \omega(B) / \omega(B= 0) = \lambda ^2 \omega \nu(E_F)  \Delta G(B) / G(B=0)$ between the typical  phonon frequency magnetic dependence  and magneto conductance, is expected to hold \cite{fuchs}.   High electron phonon coupling has already been reported in graphene  concerning in plane optical modes \cite{castroneto07}. In the present case however, strong electron phonon coupling between  transverse ZO' vibrations  and  strongly anisotropic transport in the thin graphite layer is not straight-forward  but can be understood if the electrical contacts between the two  electrodes  and the graphite foil take place through distinct graphene mono-layers which is highly probable.

In conclusion we have shown evidence of  differential conductance  sharp dips in a  suspended thin layer of graphite  with 30 graphene foils. These dips can be interpreted within a simple model of dynamical Coulomb blockade with an environment
strongly coupled to the  lowest energy optical phonon mode ZO' of graphite.
The magnetic field dependence of the effect corroborates this interpretation.

Aknowlegments: We aknowledge M. Kociak for the transmission electron microscopy pictures,  J.N.Fuchs and M.Goerbig for fruitful discussions on the electron-phonon coupling in graphite.

\end{document}